\RequirePackage{fix-cm}
\documentclass[twocolumn,epjc3]{svjour3}  
\smartqed  
\RequirePackage{graphicx}
\newcommand{\dd}{{\rm d}}
\newcommand{\dis}{{\rm DIS}_{\gamma}}
\newcommand{\lp}{\left. }
\newcommand{\rp}{\right. }

\newcommand{\lee}{\left[ }
\newcommand{\re}{\right] }
\newcommand{\lgg}{\left\{ }
\newcommand{\rg}{\right\} }
\newcommand{\ms} {\overline{\rm MS}}
\newcommand{\beq}{\begin{equation}}
\newcommand{\eeq}{\end{equation}}
\newcommand{\bea}{\begin{eqnarray}}
\newcommand{\eea}{\end{eqnarray}}
\journalname{Eur. Phys. J. C}
\begin{document}


\title{New information on photon fragmentation functions
}


\author{Michael Klasen\thanksref{e1,addr1}
        \and
        Florian K\"onig\thanksref{e2,addr1} 
}

\thankstext{e1}{e-mail: michael.klasen@uni-muenster.de}
\thankstext{e2}{e-mail: f.koenig@uni-muenster.de}


\institute{Institut f\"ur Theoretische Physik, Westf\"alische
 Wilhelms-Universit\"at M\"unster, Wilhelm-Klemm-Stra\ss{}e 9,
 D-48149 M\"unster \label{addr1}
}

\date{Received: date / Accepted: date}

\maketitle

\begin{abstract}
Thermal photons radiated in heavy-ion collisions represent an important signal
for a recently discovered new state of matter, the deconfined quark-gluon plasma.
However, a clean identification of this signal requires precise knowledge of the
prompt photons produced simultaneously in hard collisions of quarks and gluons,
mostly through their fragmentation. In this paper, we demonstrate that PHENIX data
on photons produced in proton-proton collisions with low transverse momenta allow
to extract new information on this fragmentation process. While existing data
do not yet convincingly favor one parameterization (BFG II) over the two other
frequently used photon fragmentation functions (BFG I and GRV NLO), the data sets
recorded by PHENIX and STAR at BNL RHIC in 2013 with tenfold higher statistics
should allow for such an analysis.
\keywords{Photons \and Fragmentation functions}
\PACS{12.38.Bx \and 13.85.Qk \and 14.70.Bh}
\end{abstract}

\section{Introduction}

In the early Universe, at temperatures above a critical temperature of about
$T_{\rm crit.}\simeq10^{12}$ K or 170 MeV, quarks and gluons are believed to have
existed in a new, deconfined state of matter, before they were bound by strong
interactions into protons and nuclei. Relativistic heavy-ion colliders such as
BNL RHIC and CERN LHC allow today to re-create this state, the so-called
quark-gluon plasma (QGP), on earth, albeit only for very short times of about
10$^{-23}$ s. An important signal for the presence of a QGP and
a good probe of its properties is the radiation of thermal photons with low
transverse momenta (typically $\leq 4$ GeV) from the deconfined partons
before thermalisation, in the thermal bath, during expansion and cooling of
the QGP, and finally from the thermal hadron gas \cite{Adare:2008ab,Wilde:2012wc}.

The interpretation of inclusive photon measurements is complicated by the
fact that photons are also produced in hadron (mostly neutral pion) decays,
which must be reliably
subtracted from the experimental data, as well as in hard scatterings of the quarks
and gluons in the colliding ions. At high transverse momenta, photons are mostly
produced directly, whereas in the interesting low-transverse momentum range
they stem predominantly from quark and gluon fragmentation \cite{Klasen:2013mga}.

The probability for quark and gluon fragmentation into photons can unfortunately
not be computed in perturbative QCD, but must be parameterized with photon
fragmentation functions (FFs) $D_{\gamma/q,g}(z,Q^2)$. Their dependence on the longitudinal
momentum fraction $z$ transferred from the parton to the photon is unknown and
therefore modeled at a starting scale $Q_0$. It is then evolved using QCD
renormalization group equations to higher scales $Q$, where experimental data
are available and can be used to constrain the theoretical ansatz.
 
Traditionally, these data have been taken from $e^+e^-$ colliders in order to avoid
theoretical uncertainties from the initial state and, in the absence of usable data
on prompt photons, from the production of vector mesons \cite{Abachi:1989em,%
Buskulic:1995gm}, assuming that they dominate the hadronic fluctuations into the
photon \cite{Bourhis:1997yu,Gluck:1992zx}. Today, however, the parton density functions
(PDFs) in the proton are known with much better precision than the photon FFs
\cite{Lai:2010vv}, and a wealth of new data on prompt photon production has
been taken in hadronic collisions \cite{Aurenche:2006vj}. In particular, the
PHENIX collaboration at BNL RHIC have analyzed 4 pb$^{-1}$ of 2006 $pp$ collision
data at $\sqrt{s}=200$ GeV for the production of {\em nearly} real photons with
transverse momenta in the range 1 GeV $<p_T<$ 5 GeV using a single-electron
trigger, which greatly reduced the background from light meson decays
\cite{Adare:2012vn}. These data are complemented by and overlap with real
photon data in the range $p_T>$ 4 GeV.

In this paper, we demonstrate that prompt photon data from BNL RHIC allow
in principle to extract new information on
the photon FFs. By separating the data into a control region of large transverse
momenta (above 10 GeV) and a signal region (below 5 GeV) dominated by directly
produced and fragmentation photons, respectively, we first establish the
reliability of the FF-independent parts of our perturbative QCD calculation in
the control region, before we perform chi-square tests of the three available
modern FFs (BFG I, BFG II \cite{Bourhis:1997yu} and GRV NLO \cite{Gluck:1992zx}) in
the signal region.

\section{Photon fragmentation functions}
\label{sec:2}

When a photon is radiated from a massless final-state quark, it exhibits a
collinear singularity that must be absorbed into a non-perturbative FF
$D_{\gamma/q}(z,Q^2)$. At next-to-leading order (NLO) of perturbative QCD, also
gluons fragment into photons through intermediate quarks, which gives rise to the
corresponding FF $D_{\gamma/g}(z,Q^2)$. The evolution of these FFs with the scale
$Q$ is described by renormalization group equations \cite{Koller:1978kq},
\bea
 \label{eq:frag_evol1}
 \frac{\dd D_{\gamma/q}(Q^2)}{\dd \ln Q^2} &=&
   \frac{\alpha       }{2\pi} P_{\gamma\leftarrow q}\otimes D_{\gamma/\gamma}(Q^2)
  +\frac{\alpha_s(Q^2)}{2\pi} \\
  &\times&\left[ P_{q\leftarrow q} \otimes D_{\gamma/q}(Q^2)
  +      P_{g\leftarrow q} \otimes D_{\gamma/g}(Q^2) \right] ,\nonumber\\
 \label{eq:frag_evol2}
 \frac{\dd D_{\gamma/g}(Q^2)}{\dd \ln Q^2} &=&
   \frac{\alpha       }{2\pi} P_{\gamma\leftarrow g}\otimes D_{\gamma/\gamma}(Q^2)
  +\frac{\alpha_s(Q^2)}{2\pi}\\
  &\times&\left[ P_{q\leftarrow g} \otimes D_{\gamma/q}(Q^2)
  +      P_{g\leftarrow g} \otimes D_{\gamma/g}(Q^2) \right] ,\nonumber\\
 \label{eq:frag_evol3}
 \frac{\dd D_{\gamma/\gamma}(Q^2)}{\dd \ln Q^2} &=&
   \frac{\alpha       }{2\pi} P_{\gamma\leftarrow\gamma}\otimes D_{\gamma/\gamma}(Q^2)
  +\frac{\alpha}{2\pi}\\
  &\times&\left[ P_{q\leftarrow\gamma} \otimes D_{\gamma/q}(Q^2)
  +      P_{g\leftarrow\gamma} \otimes D_{\gamma/g}(Q^2) \right], \nonumber
\eea
which are coupled through the perturbatively calculable time-like
Altarelli-Parisi splitting functions $P_{j\leftarrow i}$ \cite{Floratos:1981hs}.
Note that, contrary to the evolution equations of partons in hadrons, those
of the photon also contain inhomogeneous terms related to its pointlike
contribution.

In leading order (LO) of the electromagnetic coupling constant $\alpha$, the
third evolution equation, Eq.\ (\ref{eq:frag_evol3}),
can be directly integrated with the result $D_{\gamma/\gamma}(z,Q^2)=\delta
(1-z)$. Furthermore, in LO of the strong coupling constant $\alpha_s$,
only the evolution equation of the quark-photon FF
\beq
 \frac{\dd D_{\gamma/q}(z,Q^2)}{\dd\ln Q^2} =
 \frac{\alpha}{2\pi} P_{\gamma\leftarrow q}(z)
\eeq
survives, which can also be integrated with the result
\beq
 D_{\gamma/q}(z,Q^2)=\frac{\alpha}{2\pi}P_{\gamma\leftarrow q}(z)\ln\frac{Q^2}
 {Q_0^2}+D_{\gamma/q}(z,Q_0^2).
 \label{eq:plsolfrag}
\eeq
The first term in Eq.\ (\ref{eq:plsolfrag}) is the perturbatively calculable
pointlike solution, while the second term is a hadronic boundary condition,
which has to be fitted to experimental data.

In the modified Minimal Subtraction ($\ms$) scheme \cite{Bardeen:1978yd}, the
%
\begin{table}
\caption{\label{tab:photfrag}Current parameterizations of the photon FFs.
 $\rho$, $\omega$ and $\phi$ contributions can be added coherently or
 incoherently in Vector Meson Dominance (VMD) models.
 $N_g$ is the normalization of the gluon FF at the starting scale.}
\begin{center}
\begin{tabular}{|cc|ccccc|}
\hline
 $\!\!$Group & Set & Year & $Q_0^2$   & Factor.\ & VMD             & $\Lambda_{\ms}^{N_f=4}$      \\
       &     &      & $\!\!$(GeV$^2$)$\!\!$ & Scheme   & Model           & (MeV)\\
\hline
\hline
 $\!\!$BFG   & I   & 1998 & 2         & $\ms$    & coh., $N_g$ free\ \  & 230 \\
 $\!\!$BFG   & II  & 1998 & 2         & $\ms$    & coh., $N_g$ fixed & 230 \\
 $\!\!$GRV   & $\!\!\!$NLO$\!\!\!$ & 1993 & 0.3       & $\dis$   & incoherent         & 200 \\
\hline
\end{tabular}
\end{center}
\end{table}
%
inclusive NLO cross section for $e^+e^-\to\gamma X$ is \cite{Altarelli:1979kv}
\bea
 \frac{1}{\sigma_0} \frac{\dd\sigma(Q^2)}{\dd z} &=& \sum_q 2 e_q^2 \lgg
 D_{\gamma/q}(Q^2)
+\frac{\alpha}{2\pi} e_q^2 C_\gamma
 +\frac{\alpha_s(Q^2)}{2\pi}\rp\nonumber\\
 &\times&\lp \lee C_q \otimes D_{\gamma/q}(Q^2)
 +   C_g \otimes D_{\gamma/g}(Q^2)\re
 \rg,
\eea
where $\sigma_0=4\pi\alpha^2N_C/(3Q^2)$ is $N_C=3$ times \mbox{the cross}
section for $e^+e^-\rightarrow\mu^+\mu^-$, $e_q$ is the fractional quark charge,
the factor of two comes from
$D_{\gamma/q}(Q^2)=D_{\gamma/\overline{q}}(Q^2)$, and $C_{q,g}$ stand for the
time-like Wilson coefficients of transverse and longitudinal partonic cross
sections. In the $\dis$ factorization scheme, the singular transverse photonic
Wilson coefficient $C_{\gamma}^T\propto\ln[z^2(1-z)]$ can be absorbed into the
quark FF,
thereby increasing the perturbative stability \cite{Gluck:1992zx}.

\begin{figure}
 \centering
 \includegraphics[width=\columnwidth]{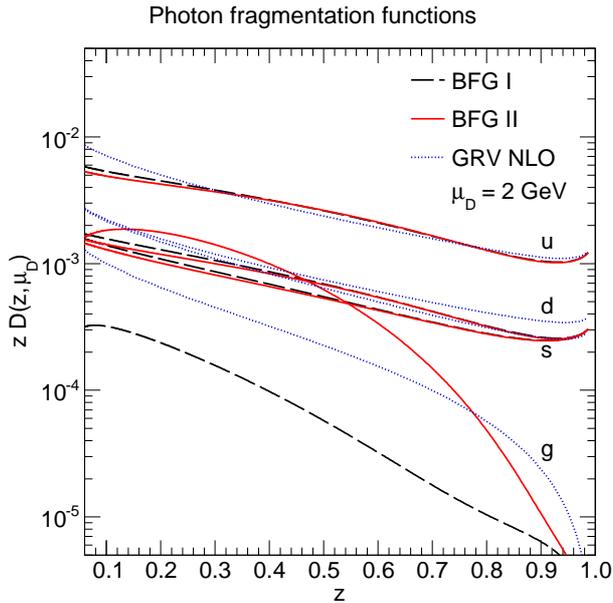}
 \caption{\label{fig:1}Quark ($u$p, $d$own, and $s$trange) and gluon ($g$)
 FFs into photons at the scale $Q=\mu_D=2$ GeV as parameterized by the BFG
 \cite{Bourhis:1997yu} and GRV collaborations \cite{Gluck:1992zx}.}
\end{figure}

The hadronic input in Eq.\ (\ref{eq:plsolfrag}), and similarly for the gluon,
can unfortunately not be determined from inclusive photon
production in $e^+e^-$ annihilation, since the experimental data are very
limited and furthermore dominated by the pointlike quark-photon FF
\cite{Ackerstaff:1997nha,GehrmannDeRidder:1998ba}. Therefore,
all current parameterizations assume vector-meson dominance (VMD) of
hadronic fluctuations into the photon to model the photon fragmentation
at low scales. The most relevant input parameters are summarized in Tab.\
\ref{tab:photfrag}.
In particular, BFG \cite{Bourhis:1997yu} work in the $\ms$ scheme and choose
a higher scale $Q_0$ and slightly larger QCD scale parameter $\Lambda$ for
$N_f=4$ flavors than GRV \cite{Gluck:1992zx}, who use the $\dis$ scheme.
Our perturbative calculation is then of course adjusted accordingly \cite{Aurenche:2006vj}.
Heavy
quarks of mass $m_h$ are included above their production thresholds with boundary
conditions $D_{\gamma/h}(z,m_h^2)=D_{\gamma/\bar{h}}(z,m_h^2)=0$.
As can be seen from Fig.\ \ref{fig:1}, these assumptions lead to
%
%
good agreement on the (mostly pointlike) quark FFs, but the gluon FFs
differ widely (by up to an order of magnitude), even among BFG I and BFG II.
The factorization scale $Q=\mu_D=2$ GeV has been chosen here in accordance
with the typical transverse momenta to be analyzed below.

\section{Subprocess contributions}
\label{sec:3}

\begin{figure}
 \centering
 \includegraphics[width=\columnwidth]{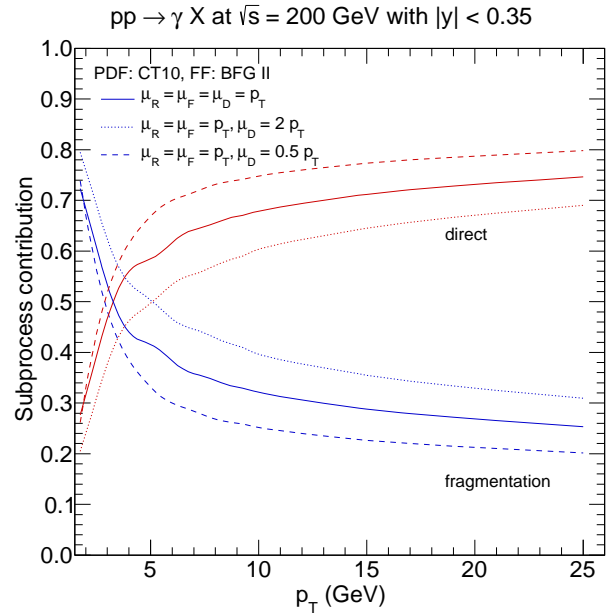}
 \caption{\label{fig:2}Fractional contributions of direct and fragmentation
 processes to inclusive photon production at BNL RHIC as a function of $p_T$
 for three different choices of the photon fragmentation scale $\mu_D$.}
\end{figure}

In proton-proton collisions, photons are not only produced by fragmentation
of the colliding quarks and gluons, but also directly in processes like
quark-antiquark fusion, $q\bar{q}\to\gamma g$, and QCD Compton scattering,
$qg\to\gamma q$. Since we want to separate the PHENIX data set into a signal
and a control region, dominated by fragmentation and direct production,
respectively, we must first establish the corresponding $p_T$ regions. To
this end, we compute the fractional subprocess contributions assuming a fixed
set of parton densities given by the CT10 parameterization  \cite{Lai:2010vv},
which are well constrained in the region of $x_T=2p_T/\sqrt{s}=0.01-0.1$ relevant here,
and identifying the renormalization scale $\mu_R$, the proton factorization scale
$\mu_F$ and the photon fragmentation scale $\mu_D$ with the central hard scale
of the process, the photon transverse momentum $p_T$. Fig.\ \ref{fig:2} then
shows that fragmentation processes dominate for $p_T\leq5$ GeV in $pp$ collisions
at $\sqrt{s}=200$ GeV, while for $p_T>10$ GeV direct processes account for $60-75$\%
of the total cross section, depending on $\mu_D$. If one wants to fix the
fragmentation-independent parts of the NLO QCD calculation \cite{Aurenche:2006vj}, it
is therefore preferable to choose $\mu_D=0.5\,p_T$ in order to minimize the fragmentation
contribution.

%
%

\section{Comparison with PHENIX data}
\label{sec:4}

\begin{figure}
 \centering
 \includegraphics[width=\columnwidth]{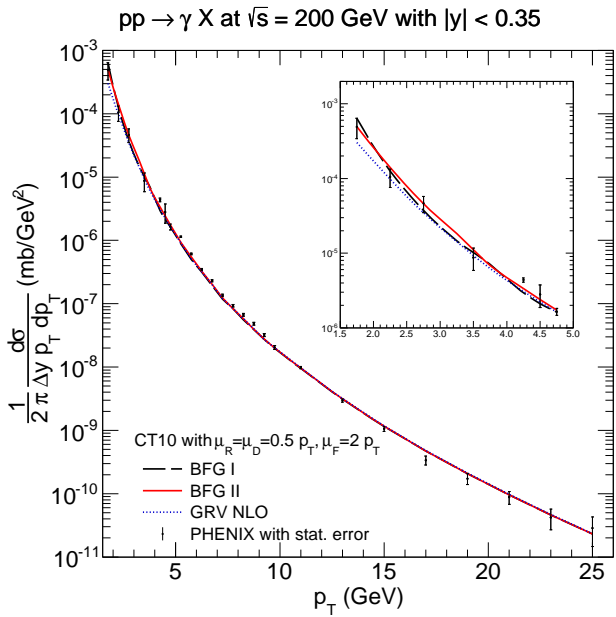}
 \caption{\label{fig:3}Transverse-momentum distribution of inclusive
 photons as predicted by three different FFs and compared to PHENIX data
 with statistical errors only at low (insert) and high $p_T$ \cite{Adare:2012vn}.}
\end{figure}

Having fixed our signal and control regions as described above, we next
allow all three scales to vary independently among the choices $(0.5;1;2)\, p_T$
in the control region ($p_T>10$
GeV) and fit them to the PHENIX data, using geometrical binning and
statistical errors only, as the systemtatic errors are dominated by hadron
decay uncertainties and largely correlated among different $p_T$-bins \cite{Adare:2012vn}.
We find a mimimal value of $\chi^2$/d.o.f.
of 1.2 for the combination $\mu_R=\mu_D=0.5\,p_T$ and $\mu_F=2\,p_T$ for the BFG I
and II FFs and somewhat larger for GRV NLO, which is in good accordance with
our observation above that $\mu_D=0.5\,p_T$ should be preferred.
Although higher-order QCD corrections are of course in principle important,
in particular at low $p_T$, they can be subsumized by an appropriate choice
of scale. We have exploited this freedom by normalizing the theory to the
data, in this way effectively fitting the higher-order terms.
Note also that when $\mu_D$ falls below the starting scale $Q_0=\sqrt{2}$ GeV,
numerical results from the BFG parameterizations of the FFs are no longer available
and $\mu_D$ must at least be frozen there. In order to avoid the appearance
of large logarithms (like $\log\mu_R/\mu_D$), we have chosen to freeze all three scales
($\mu_R$, $\mu_F$ and $\mu_D$) at $Q_0$ in  the short- and long-distance
parts of our calulation. The error committed in this way is then at least of
next-to-next-to-leading order, coming only from the uncompensated parts
in the PDF and FF evolutions, and it affects all three FFs in a similar 
and only logarithmic way, ensuring a subdominant impact on our comparison with data.
The goodness of our fit and its
independence of the choice of FF can also be observed in the high-$p_T$ region of
Fig.\ \ref{fig:3}.
%
%

We can then perform a $\chi^2$ test of the three different FFs in the
signal region ($p_T<5$ GeV, see insert of Fig.\ \ref{fig:3}),
finding an acceptable minimal value of $\chi^2$/d.o.f.\ of 2.8 for BFG II, while
the BFG I and GRV NLO hypotheses lead to significantly larger values of
5.2 and 4.5, respectively, and can be rejected at a confidence level of 99\%.
Looking at Fig.\ \ref{fig:3}, these values of $\chi^2$/d.o.f.\
are obviously dominated by the exceptionally high point at $p_T=4.25$ GeV,
which together with the point at $p_T=4.75$ GeV comes from the real photon analysis.
Although the other data points from the nearly real photon analysis overlap with these
two real photon data points within their
respective $p_T$-correlated systematic errors (see Fig.\ 2 of Ref.\
\cite{Adare:2012vn}), the systematic errors differ among the two analyses.
If we omit the two real photon data points from the fit, we then find values
of $\chi^2$/d.o.f. of 0.68 for BFG II, 0.61 for BFG I and 0.63 for GRV.
The current level of statistical (nearly real photons) and systematic
(real photons) precision thus does not yet allow to obtain stringent information
on the photon FF. An improvement of about a factor of five in the
statistical error would still be needed to apply our method successfully.

\section{Conclusions}
\label{sec:4}

In this paper, we have seen that the combined virtual and real
photon data from PHENIX seem to favor the BFG II parameterization with its relatively
large gluon distribution over BFG I and GRV. This observation is, however, driven by an
exceptionally high real-photon data point, which overlaps with the virtual
photon data only within its large systematic error. The published virtual photon
data from PHENIX alone do not yet allow for a conclusive distinction of
the three available photon FFs and would require a reduction in their
statistical error of at least a factor of five.

In the absence of new $e^+e^-$ data, e.g.\ from a Linear Collider,
our study shows nevertheless the potential of future inclusive photon measurements
at BNL RHIC and CERN LHC to constrain the photon FFs with hadron collider
data. In fact, much higher luminosities of 574 and 526 pb$^{-1}$ have already
been recorded in 2013 by PHENIX and STAR, respectively, in $pp$ collisions at BNL
RHIC and 5$-$10 pb$^{-1}$ by the ALICE experiment at CERN LHC with $\sqrt{s}=7-8$ TeV.
Unfortunately, at the LHC
limitations of band width impede to trigger on low-$p_T$ data. For the
suppression of meson decays, it seems crucial
to exploit new experimental techniques such as electron triggers for
nearly real photon detection.

In the future it might be possible to also exploit photon-jet correlations
at BNL RHIC \cite{Belghobsi:2009hx}. Indeed, photon-hadron correlations
have already been studied, and the component of the photon momentum
perpendicular to a trigger hadron has been extracted \cite{Hanks:2009fj}.
For decay and fragmentation photons, it was shown to be with about 0.5 GeV
significantly smaller than the one for directly produced photons ($\sim$ 0.8 GeV).

\begin{acknowledgements}

We thank A.\ Gehrmann-De Ridder and Y.\ Yamaguchi for providing us with the FF
routines and detailed information on the PHENIX data, respectively. We also thank
C.\ Klein-B\"osing and J.\ Wessels for useful comments on the manuscript.

\end{acknowledgements}


\end{document}